\documentclass[aip,jcp,12pt,reprint,floatfix]{revtex4-1}

\usepackage{graphicx}
\usepackage[version=3]{mhchem}

\begin{document}

\title{Excited-State Dynamics in SO$_2$: II. The Role of Triplet States in the Bound State Relaxation Studied by Surface-Hopping Simulations} 

\author{Sebastian Mai}
\affiliation{Institute of Theoretical Chemistry, University of Vienna, W\"ahringer Str. 17, 1090 Vienna, Austria}

\author{Philipp Marquetand}
\email[]{philipp.marquetand@univie.ac.at}
\affiliation{Institute of Theoretical Chemistry, University of Vienna, W\"ahringer Str. 17, 1090 Vienna, Austria}

\author{Leticia Gonz\'alez}
\affiliation{Institute of Theoretical Chemistry, University of Vienna, W\"ahringer Str. 17, 1090 Vienna, Austria}

\date{\today}

\begin{abstract}
The importance of triplet states in the photorelaxation dynamics of \ce{SO2} is studied by mixed quantum-classical dynamics simulations. Using the Surface Hopping including ARbitrary Couplings (\textsc{Sharc}) method, intersystem crossing processes caused by spin-orbit coupling are found occuring on an ultrafast time scale (few 100 fs) and thus competing with internal conversion. While in the singlet-only dynamics only oscillatory population transfer between the $^1B_1$ and $^1A_2$ states is observed, in the dynamics including singlet and triplet states we find additionally continuous ISC to the $^3B_2$ state and to a smaller extent to the $^3B_1/^3A_2$ coupled states. The populations obtained from the dynamics are discussed with respect to the overall nuclear motion and in the light of recent TRPEPICO studies \lbrack Wilkinson et al., paper I\rbrack.
\end{abstract}



\maketitle 


\section{Introduction}\label{sec:introduction}

During the last decades, significant progress has been made improving multi-reference ab initio techniques\cite{gonzalez_2012_C_28} and ultra-fast, time-resolved spectroscopic methods.\cite{hertel_2006_RPP_1897, quack_2011__} Consequently, the interest in small model systems is revived because they can be studied in great detail. One prominent example is the \ce{SO2} molecule, owing to its relevance in atmospheric processes and the fact that its low-energy excited-state dynamics is still not completely understood.

In a companion paper,\cite{wilkinson_2013_JCP_} (hereafter referred to as Paper~I) the experimental and theoretical efforts of the last 80~years to characterize the electronic excited states of \ce{SO2} have been surveyed in detail, thus, here only a short summary is provided.
Already in the 1930s, absorption spectra of \ce{SO2} were recorded.\cite{watson_1931_PR_1484,clements_1935_PR_224} 
The absorption spectrum features three prominent bands in the near- and medium-UV range, named the forbidden band and the first and second allowed bands.\cite{heicklen_1980_RCI_315} Especially the first allowed band between 3.6~eV and 5.1~eV with its intricate vibrational structure was subject to numerous analyses.\cite{metropolis_1941_PR_295, hamada_1974_CJP_1443, hamada_1975_CJP_2555, shaw_1976_CP_155, shaw_1976_CP_165, kullmer_1985_CP_423, baskin_1995_CP_181, bae_1997_CPL_385} A large number of experimental works was reviewed by Herzberg\cite{herzberg_1966__} and Heicklen et al.\cite{heicklen_1980_RCI_315}

From the point of view of modern ab initio quantum chemistry,\cite{hillier_1971_MP_193, kamiya_1991_BCSJ_2792, muller_1994_CP_107, katagiri_1997_JMS_589, xie_2000_CPL_503, koppel_2000_IJQC_942, li_2006_SCSB_289, dehestani_2011_CTC_1, leveque_2013_JCP_11, xie_2013_JCP_14305} the vibrational band system of the first allowed band arises from the transition from the ground state to the vibronically coupled $1^1B_1/1^1A_2$ system (the two lowest excited singlet states). M\"uller and K\"oppel\cite{muller_1994_CP_107} treated this two-state-system using full-dimensional quantum dynamics (QD) and found that the system remains primarily on the lower adiabatic potential energy surface (PES) after excitation. Later, L\'ev\^eque et al.\cite{leveque_2013_JCP_11} extended these QD studies, using accurate MRCI potentials, and were able to semi-quantitatively reproduce the lower energy part of the first allowed band and obtained valuable information about the interactions which give rise to the structure of the spectrum.

Furthermore, there is significant experimental evidence\cite{kusch_1939_PR_850, douglas_1958_CJP_147, brand_1976_CJP_186, kullmer_1985_JCP_2712, watanabe_1983_FD_365, watanabe_1983_JPC_906} 
that the excited state dynamics within the first allowed band system is also affected by the presence of triplet states. \ce{SO2} exhibits in the region between 3.1~eV and 3.6~eV a weak absorption profile (the forbidden band), which was shown to arise from excitation to triplet states by Douglas\cite{douglas_1958_CJP_147} by means of the Zeeman effect. 
Considerable experimental effort had been devoted to identify the number and character of the low-lying triplet states. While the origin and symmetry of the $\tilde{a}^3B_1$ state\cite{merer_1963_DFS_127, brand_1971_JMS_616, brand_1973_JMS_404, hallin_1976_CJP_2118, vikesland_1974_JCP_660, baskin_1995_CP_181, joens_1996_CPL_659, huang_2000_JMS_151, huang_2001_JCP_1187} and the $\tilde{b}^3A_2$ state\cite{merer_1963_DFS_127, brand_1973_JMS_404, huang_2001_JCP_1187} have been well established, the location of the $\tilde{c}^3B_2$ state\cite{brand_1973_JMS_404} has yet to be determined experimentally. The experimental findings were confirmed and further elucidated upon by means of modern ab initio methods, showing that these three triplet states are present at energies slightly lower than the corresponding singlet states. 
Consequently, intersystem crossing (ISC) between singlet and triplet states is plausible and the photodynamics of \ce{SO2} might be influenced by both ISC and internal conversion (IC) processes. Xie et al.\cite{xie_2013_JCP_14305} performed full-dimensional QD calculations on the $1^1B_1/1^1A_2$ system and also included the $^3B_1$ triplet state. They found a substantial and ultrafast population transfer from $^1A_2$ to the triplet state. The most recent and comprehensive experimental study on \ce{SO2} excited-state dynamics is based on time-resolved photoelectron photoion coincidence spectroscopy (TRPEPICO) and is reported in Paper~I. There, ISC time constants ranging from 150 to 750 fs have been measured.

The present paper is an attempt to unravel the excited state dynamics of \ce{SO2} theoretically, for the first time including all singlet and triplet states in the relevant energy range. To this aim, ab initio surface hopping molecular dynamics (AIMD) is employed using the \textsc{Sharc} code.\cite{richter_2011_JCTC_1253} \textsc{Sharc} (Surface Hopping including ARbitrary Couplings) can treat IC and ISC, mediated by non-adibatic couplings and spin-orbit couplings (SOC), respectively, on the same footing. 
This is achieved by performing surface hopping in a basis of spin-orbit-coupled electronic states, which are obtained by diagonalization of the potential energy matrix including the spin-orbit couplings. 
Other applications of \textsc{Sharc} can be found in Refs.~\onlinecite{richter_2011_JCTC_1253, marquetand_2011_FD_261, bajo_2011_JPCA_2800,richter_2012_JPCL_3090, mai_2013_C_2920}.
Here, the simulations are focused on the energy range corresponding to the forbidden and first allowed band of the experimental \ce{SO2} absorption spectrum.\cite{golomb_1962_JCP_958} In order to validate our methodology, we also performed dynamics simulations in the singlet manifold, giving the possibility to compare the outcome to the QD results of L\'ev\^eque et al.\cite{leveque_2013_JCP_11}. 


\section{Methodology}\label{sec:methodology}


\subsection{Surface hopping including arbitrary couplings}\label{ssec:SHARC}

Surface hopping dynamics according to Tully's fewest switches criterion\cite{tully_1990_JCP_1061} is usually performed on the PESs of the eigenfunctions of the molecular Coulomb Hamiltonian (MCH). This Hamiltonian is the sum of the electronic kinetic energy and the potential energy arising from the Coulomb interaction of the electrons and nuclei with each other, i.e.,
\begin{equation}
  \hat{H}_{\text{el}}^{\text{MCH}}
  =\hat{K}_{\text{e}}
  +\hat{V}_{\text{ee}}
  +\hat{V}_{\text{ne}}
  +\hat{V}_{\text{nn}}.
\end{equation}
Ab initio quantum chemistry software mostly obtains electronic wavefunctions as eigenfunctions of this operator and properties like gradients and non-adiabatic couplings are available for these wavefunctions, making non-adiabatic dynamics possible. However, in order to account for phenomena like light-matter interaction or ISC additional terms in the Hamiltonian are necessary, e.g. dipole couplings or spin-orbit couplings:%
\begin{equation}
  \hat{H}_{\text{el}}^{\text{total}}
  =\hat{H}_{\text{el}}^{\text{MCH}}
  +\hat{H}_{\text{el}}^{\text{coup}}.
\end{equation}

Because of the classical approximations inherent to surface-hopping, nuclear motion has to follow the potential energy surfaces (PESs) of the eigenfunctions of the total electronic Hamiltonian $\hat{H}_{\text{el}}^{\text{total}}$. Henceforth, the basis of the eigenfunctions of $\hat{H}_{\text{el}}^{\text{total}}$ will be referred to as the diagonal basis (superscript diag), since all off-diagonal potential couplings are transformed away in this basis. Although wavefunctions of the total Hamiltonian $\hat{H}_{\text{el}}^{\text{total}}$ are available in few cases,\cite{yabushita_1999_JPCA_5791} gradients and non-adiabatic couplings cannot currently be obtained. 
Instead, in standard quantum chemistry packages, the PESs are usually obtained in the basis of the $\hat{H}_{\text{el}}^{\text{MCH}}$ eigenfunctions ($|\phi_\alpha^{\text{MCH}}\rangle$). Thus, in the recently developed \textsc{Sharc} 
methodology\cite{richter_2011_JCTC_1253} we define a model space including the $N$ lowest eigenfunctions $|\phi_\alpha^{\text{MCH}}\rangle$. The electronic wavefunction $|\Psi_{\text{el}}\rangle$ is expanded in the basis of these electronic wavefunctions:
\begin{equation}
  |\Psi_{\text{el}}\rangle=\sum_\alpha^N|\phi_\alpha^{\text{MCH}}\rangle c_\alpha^{\text{MCH}}.
  \label{eq:wavefunction}
\end{equation}
The total electronic Hamiltonian is represented in the MCH basis by the matrix $\mathbf{H}^{\mathrm{MCH}}$ with elements
\begin{equation}
  H^{\mathrm{MCH}}_{\beta\alpha}=\langle\phi_\beta^{\mathrm{MCH}}|\hat{H}_{\text{el}}^{\text{total}}|\phi_\alpha^{\mathrm{MCH}}\rangle
\end{equation}
and in the diagonal basis by the matrix $\mathbf{H}^{\mathrm{diag}}$, where
\begin{equation}
  H^{\mathrm{diag}}_{\beta\alpha}=\langle\phi_\beta^{\mathrm{diag}}|\hat{H}_{\text{el}}^{\text{total}}|\phi_\alpha^{\mathrm{diag}}\rangle
\end{equation}
The matrix representations of the total electronic Hamiltonian are related by a unitary transformation:
\begin{equation}
  \mathbf{U}^\dagger\mathbf{H}^{\text{MCH}}\mathbf{U}=\mathbf{H}^{\text{diag}}.
\end{equation}

Inserting the wavefunction ansatz \eqref{eq:wavefunction} into the time-dependent Schr\"odinger equation leads to the differential equation governing the evolution of the wavefunction coefficients:%
\begin{equation}
  \frac{\partial}{\partial t}\mathbf{c}^{\text{MCH}}=
  -\left[
    \frac{\text{i}}{\hbar}\mathbf{H}^{\text{MCH}}
    +\mathbf{K}^{\text{MCH}}
  \right]
  \mathbf{c}^{\text{MCH}},
  \label{eq:evolution}
\end{equation}
or, equivalently:
\begin{equation}
  \frac{\partial}{\partial t}\mathbf{c}^{\text{diag}}=
  -\mathbf{U}^\dagger\left[
    \frac{\text{i}}{\hbar}\mathbf{H}^{\text{MCH}}
    +\mathbf{K}^{\text{MCH}}
  \right]\mathbf{U}
  \mathbf{c}^{\text{diag}},
  \label{eq:evolution2}
\end{equation}
where the elements of $\mathbf{K}^{\text{MCH}}$ are the non-adiabatic couplings $\langle\phi_\beta^{\text{MCH}}|\partial/\partial t|\phi_\alpha^{\text{MCH}}\rangle$. 

The integration of equation~\eqref{eq:evolution2} is performed here using the very stable \textit{local diabatization} algorithm proposed by Granucci et al.,\cite{granucci_2001_JCP_10608} which yields the following equation for the propagation of the coefficients:
\begin{equation}
  \mathbf{c}^{\text{diag}}(t)=
  \underbrace{
    \mathbf{U}^\dagger(t)
    \mathbf{S}^\dagger
    \text{e}^{
      -\frac{\text{i}}{\hbar}\left[
        \mathbf{H}^{\text{MCH}}(t_0)
        +\mathbf{S}\mathbf{H}^{\text{MCH}}(t)\mathbf{S}^\dagger
      \right]
      \frac{\Delta t}{2}
    }
    \mathbf{U}(t_0)
  }_{
    \mathbf{A}^{\text{diag}}(t,t_0)
  }
  \mathbf{c}^{\text{diag}}(t_0).
  \label{eq:finalprop}
\end{equation}
The overlap matrix $\mathbf{S}$ with matrix elements $S_{\beta\alpha}=\langle\phi_\beta^{\text{MCH}}(t_0)|\phi_\alpha^{\text{MCH}}(t)\rangle$ gives the unitary transformation between the bases $\{\phi^{\text{MCH}}(t_0)\}$ and $\{\phi^{\text{MCH}}(t)\}$. 

The action of the total propagator $\mathbf{A}^{\text{diag}}(t,t_0)$ on the coefficients $\mathbf{c}^{\text{diag}}$ finally describes the population transfer between the states in the diagonal basis. The surface hopping probabilities can then be calculated as:
\begin{multline}
  P_{\beta\rightarrow\alpha}=
  \left(
    1-
    \frac{
      \left|
        c_\beta^{\text{diag}}(t)
      \right|^2
    }{
      \left|
        c_\beta^{\text{diag}}(t_0)
      \right|^2
    }\right)
    \\
    \times
    \frac{
      \Re\left[
        c^{\text{diag}}_\alpha(t)
        A^*_{\alpha\beta}
        \left(
          c^{\text{diag}}_\beta(t_0)
        \right)^*
      \right]
    }{
      \left|
        c^{\text{diag}}_\beta(t_0)
      \right|^2
      -\Re\left[
        c^{\text{diag}}_\beta(t)
        A^*_{\beta\beta}
        \left(
          c^{\text{diag}}_\beta(t_0)
        \right)^*
      \right]
    }.
\end{multline}
This is a modification of the equation derived by Granucci et al.\cite{granucci_2001_JCP_10608} used in the local diabatization algorithm\cite{plasser_2012_JCP_13} available in~\textsc{Newton-X}.\cite{barbatti_2011__} We also apply decoherence as proposed in Ref.~\onlinecite{granucci_2007_JCP_11} to the diagonal coefficients.

Besides high numerical stability, the overlap matrix $\mathbf{S}$ needed for the wavefunction propagation also allows to transform the coefficients into a diabatic reference basis, which can be chosen to coincide with the ``spectroscopic'' states (e.g. $\pi\pi^*$ and $n\pi^*$). The transformation of the diagonal coefficients at time $t_i$ into the spectroscopic basis can be written as:
\begin{equation}
  \mathbf{c}^{\text{spec}}(t_i)=
  \mathbf{S}(\text{spec},0)
  \prod_{j=1}^{i}
    \mathbf{S}(t_{j-1},t_{j})
  \mathbf{U}(t_i)
  \mathbf{c}^{\text{diag}}(t_i).
\end{equation}
Here, $\mathbf{S}(t_{j-1},t_{j})$ is the matrix describing the transformation from the basis $\{\phi^{\text{MCH}}(t_j)\}$ to the basis $\{\phi^{\text{MCH}}(t_{j-1})\}$. 

In summary, molecular properties like energies, gradients and couplings are obtained in the MCH basis using ab initio quantum chemistry, within \textsc{Sharc} transformed into the diagonal basis during the surface hopping simulation and transformed back into the MCH basis or the spectroscopic basis for analysis. The electronic wavefunction propagation is carried out in a locally diabatic representation as described in Ref.~\onlinecite{granucci_2001_JCP_10608}.


\subsection{Ab initio methods and dynamics}\label{ssec:abinitio}

The on-the-fly electronic structure calculations required for the MD simulations were performed with a development version of \textsc{Columbus 7.0}.\cite{mai___} This code allows to obtain the Hamiltonian matrix including the SOC in the MCH basis, as well as gradients and wavefunction overlaps, which are needed for the local diabatization propagation, at the MRCI level of theory. The electronic wavefunction is based on orbitals coming from a CASSCF(12,9)/ano-rcc-vdzp\cite{roos_2004_JPCA_2851} calculation, state-averaged over the four lowest singlet states and three lowest triplet states (the $1^1A_1$, $1^1B_1$, $1^1A_2$ $1^1B_2$, $1^3B_1$, $1^3A_2$ and $1^3B_2$ states at the equilibrium geometry). 

Since the CASSCF(12,9) wavefunction is insufficient to correctly describe the PESs, the wavefunction was subsequently correlated with MR-CIS based on a CAS(6,6) reference space. Preliminary calculations showed that this correlation treatment is computationally very efficient and already achieves fairly accurate PESs. 
The spin-orbit matrix elements were calculated using the effective Fock-type spin-orbit operator as implemented in \textsc{Columbus} SO-CI.\cite{yabushita_1999_JPCA_5791,mai___}

\begin{figure}
  \includegraphics[scale=1]{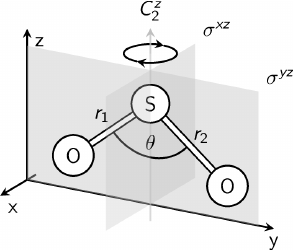}
  \caption{Geometry of the molecule, internal coordinates and symmetry elements of \ce{SO2}. The experimental values are $r_1$=$r_2$=1.432\AA\ and $\theta$=119.5$^\circ$.\label{fig:Geometry}}
\end{figure}

Throughout this paper, the nuclear motion of \ce{SO2} is discussed in terms of a set of internal coordinates consisting of the average bond length, $r_{\mathrm{sym}}$, one half of the bond length difference, $r_{\mathrm{asym}}$, and the bond angle, $\theta$:%
\begin{equation}
  r_{\mathrm{sym}}=\frac{1}{2}(r_1+r_2),\qquad
  r_{\mathrm{asym}}=\frac{1}{2}(r_1-r_2),
\end{equation}
where $r_1$, $r_2$ and $\theta$ are defined in Figure~\ref{fig:Geometry}. The MD calculations themselves were carried out in cartesian coordinates.

For the generation of the initial conditions, the ground state equilibrium geometry was optimized using the above-described MR-CIS method and harmonic frequencies were obtained at the same level of theory. The results are given in Table~\ref{tab:geom}, showing a good agreement with experimental values.
From the frequencies, a quantum harmonic oscillator Wigner distribution\cite{schinke_1995__,dahl_1988_JCP_4535} was calculated and 1000 geometries were sampled from the distribution. An absorption spectrum was simulated from single-point calculations at all 1000 geometries and initial conditions for the dynamics were selected based on the oscillator strength between ground state and excited state.\cite{barbatti_2007_JPPA_228,barbatti_2011__} Only excitations in the energy range from 4.1 to 4.6~eV were allowed. This energy range was chosen to comprise the three excitation windows employed in the experiments of paper~I, considering the energy shift of the simulated spectrum (see below, figure~\ref{fig:Spectrum}).

\begin{table}
  \centering
  \caption{Equilibrium geometry parameters and ground state vibrational frequencies from MRCIS. }
  \label{tab:geom}
  \begin{tabular}{lcccc}
    \hline
    Coordinate &This work &Xie et al.\cite{xie_2013_JCP_14305} &Tokue et al.\cite{tokue_2010_JCP_24301} &Experiment\\
    \hline
    $r_1$ (\AA)                     &1.453  &1.454   &1.455   &1.432\\
    $r_2$ (\AA)                     &1.453  &1.454   &1.455   &1.432\\
    $\theta$ ($^\circ$)             &120.5   &119.2   &118.5   &119.5\\
    \hline
    $\nu_{\text{Bend}}$ (cm$^{-1}$) &518.75  &516.22  &565.12   &517.69\\
    $\nu_{\text{sym}}$ (cm$^{-1}$)  &1165.17 &1113.52 &1115.05  &1151.38\\
    $\nu_{\text{asym}}$ (cm$^{-1}$) &1405.24 &1325.07 &1305.73  &1361.76\\
    \hline
  \end{tabular}
\end{table}

The dynamics simulations were carried out based on these initial conditions with the above-mentioned level of theory, including the four lowest singlet states and three lowest triplet states. The number of states is chosen so that the $^1A_2$ and $^1B_1$ states are included in the simulation for all time steps, making the transformation into the spectroscopic basis (see above) possible. The timestep of the integration of the nuclear motion was 0.5~fs, which is sufficient given that \ce{SO2} does not feature hydrogen atoms. The electronic wavefunction was integrated according to equation~\ref{eq:finalprop} with a timestep of 0.02~fs. The total propagation time was 700~fs.


\section{Results}\label{sec:results}

\subsection{Spectrum}\label{ssec:spectrum}

\begin{table}[b]
  \centering
  \caption{Vertical excitation energies (eV) of the singlet and triplet states included in the dynamics simulations. }
  \label{tab:vertexc}
  \begin{tabular}{lcccccc}
    \hline
    State       &This work    &MRCI\cite{xie_2013_JCP_14305}    &MRCI\cite{leveque_2013_JCP_11}     &CCSD\cite{leveque_2013_JCP_11}     &MRCI\cite{katagiri_1997_JMS_589}       &MRCI\cite{elliott_2005_JPCA_11304}\\
    \hline
    $^1B_1$     &4.46   &4.19   &4.23   &4.39   &4.15   &4.47\\
    $^1A_2$     &4.85   &4.61   &4.61   &4.84   &4.59   &4.78\\
    $^1B_2$     &6.81   &--     &--     &--     &6.34   &6.58\\
    \hline
    $^3B_1$     &3.65   &3.33   &--     &--     &3.27   &3.57\\
    $^3B_2$     &4.48   &--     &--     &--     &4.29   &4.41\\
    $^3A_2$     &4.63   &4.37   &--     &--     &4.59   &4.55\\
    \hline
  \end{tabular}
\end{table}

In Table~\ref{tab:vertexc}, we present excitation energies at the $S_0$ equilibrium geometry as obtained with MR-CIS/ano-rcc-vdzp (see section~\ref{ssec:abinitio}). The excitation energies are systematically larger than the values obtained by Xie et al.\cite{xie_2013_JCP_14305}, L\'ev\^eque et al.\cite{leveque_2013_JCP_11} and Katagiri et al.\cite{katagiri_1997_JMS_589}, but they are very close to the results of Elliott et al.\cite{elliott_2005_JPCA_11304}. Contrary to Xie et al.\cite{xie_2013_JCP_14305}, we do not find the $^3B_2$ state higher in energy than the other triplet states.

\begin{figure}
  \includegraphics[width=0.45\textwidth]{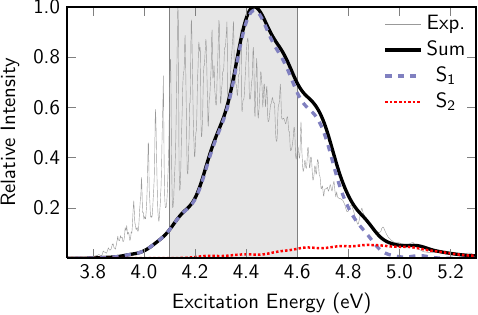}%
  \caption{Absorption spectrum arising from excitation to $S_1$ and $S_2$. The experimental spectrum of Golomb et al.\cite{golomb_1962_JCP_958} is given as well. The light-grey region from 4.1 eV to 4.6 eV indicates the excitation window for the initial condition generation.\label{fig:Spectrum}}%
\end{figure}

From the excitation energies and oscillator strengths of the 1000 initial geometries an absorption spectrum was calculated using a Gaussian convolution. The spectrum is presented in Figure~\ref{fig:Spectrum} together with the experimental spectrum.\cite{golomb_1962_JCP_958} 
The spectrum is largely dominated by excitation to the $S_1$ state, especially in the excitation window -- given by the grey area in Figure~\ref{fig:Spectrum}. The $S_2$ on the other hand is almost dark and only shows some intensity in the high-energy part of the spectrum.
Naturally, the vibrational structure of the spectrum cannot be reproduced by this semi-classical method. However, our semi-classical simulation predicts fairly accurately the total width of the spectrum (see Figure~\ref{fig:Spectrum}), with energies ranging from 3.8~eV to 5.2~eV. It is interesting to note that the simulated spectrum resembles the underlying quasi-continuum of the absorption band.

\subsection{Potential energy surfaces}\label{ssec:PES}

As already mentioned in the methodology section, within \textsc{Sharc} a number of different representations are used. Thus, a clarification of the state labelling is on order here. Even though the diagonal representation is the most appropriate basis for surface hopping dynamics, it is rather inconvenient for discussing the dynamics. Instead, we employ the MCH states for the analysis of the dynamics, since these states are not spin-mixed. Henceforth, we will refer to these states by the labels $S_n$ and $T_n$, for singlet and triplet states, respectively. Additionally, we transform the state populations into the spectroscopic basis, which very much resemble the $C_{2v}$ states. Thus, we will refer to these states by their symmetry labels, e.g. $^1A_1$ or $^3B_2$.

\begin{figure}
  \includegraphics[scale=1]{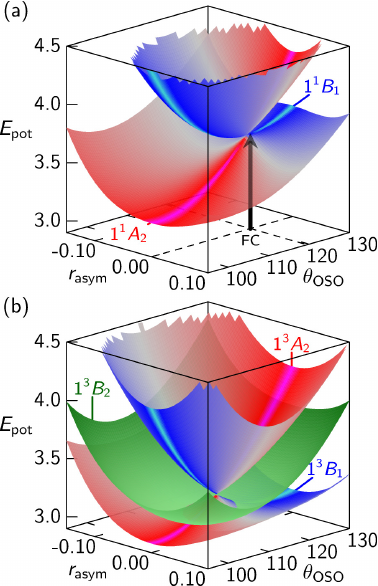}%
  \caption{Potential energy surfaces (MCH representation) surrounding the conical intersection between the singlet states $S_1$ and $S_2$ (a) and between the triplet states $T_1$, $T_2$ and $T_3$ (b) for $r_{\mathrm{sym}}$=1.5\AA. Colors indicate the spectroscopic character: red corresponds to $A_2$, blue to $B_1$ and green to $B_2$. For $r_{\text{asym}}\neq0$, $A_2$ and $B_1$ mix, which is indicated in grey. In (a), the arrow denotes the Franck-Condon point.\label{fig:PES_States}}%
\end{figure}

Figure~\ref{fig:PES_States} shows qualitative PESs of the $^1B_1$ and $^1A_2$ singlet states in (a) and of the $^3B_1$, $^3A_2$ and $^3B_2$ triplet states in (b) for $r_{\mathrm{sym}}$=1.5\AA. In (a) the singlet states show a conical intersection (CoIn) at $\theta\approx117^\circ$ and $r_{\text{asym}}=0$. It also shows the Franck-Condon (FC) point at $\theta\approx119.5^\circ$ and $r_{\text{asym}}=0$. For angles $\theta$ larger than $117^\circ$, the lower singlet surface $S_1$ corresponds to the $^1B_1$ state and $S_2$ to $^1A_2$, while for smaller angles this corresponce is reversed. For $r_{\text{asym}}\neq0$, the point group of the molecule descends to $C_s$ and $^1B_1$ and $^1A_2$ become two states of $A^{\prime\prime}$ symmetry, which avoid each other. 
In panel (b), the triplet states $^3B_1$ and $^3A_2$ show at $\theta\approx111^\circ$ a CoIn similar to the singlet case. Additionally, the $^3B_2$ state intersects the other triplet states. Since even in $C_s$ symmetry this state does not mix with the other triplet states, it retains its wavefunction character for the whole surface. 

\begin{figure}
  \includegraphics[scale=1]{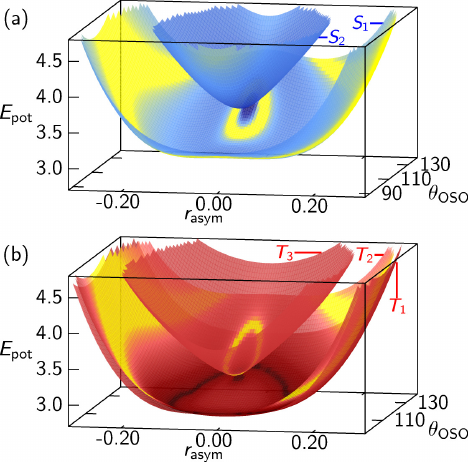}%
  \caption{Potential energy surfaces (MCH representation) of the lowest excited singlet (a) and triplet (b) surfaces for $r_{\mathrm{sym}}$=1.7~\AA. Dark shades indicate regions of strong non-adiabatic coupling. Singlet-triplet interaction regions (energy difference $<$0.05~eV$\approx$400~cm$^{-1}$) are indicated in yellow.\label{fig:PES_all}}%
\end{figure}

Some additional features of the PESs are visible in Figure~\ref{fig:PES_all}, plotted for $r_{\mathrm{sym}}$=1.7~\AA.
In panel (a), first note the double-miminum form of the lower singlet surface along the asymmetric stretch coordinate. This feature of the PES is called a pseudo-Jahn-Teller instability and is well known in the \ce{SO2} system.\cite{davidson_1983_JPC_4783}
Additionally, Figure~\ref{fig:PES_all} shows regions with a small singlet-triplet gap in yellow. The first of these regions circularly surrounds the singlet CoIn and is quite narrow. Here, the $S_1$ crosses with the upper $^3B_1/^3A_2$ surface. A much more extended region with a small singlet-triplet gap can be found for large absolute values of $r_{\mathrm{asym}}$, where the singlet and triplet surfaces become degenerate at the \ce{O-SO} dissociation limit. The triplet state closest to the singlet surface is the $^3B_2$ in this case.
Figure~4 (b) shows the triplet PESs, with the same regions of singlet-triplet interaction. Like in the singlet state case, a double-minimum shape of the two lower triplet surfaces can be observed. Note also the intersection seam (in black) of the $^3B_2$ ($^3A^\prime$) state with the lower $^3B_1/^3A_2$ ($^3A^{\prime\prime}$) surface and the near-parallel surfaces of the two lower triplet states for large values of $r_{\mathrm{asym}}$.

In order to examine the accuracy of the PESs obtained with our MR-CIS method (see section~\ref{ssec:abinitio}), we determined minimum energies and their geometries for all singlet and triplet states under consideration, as well as the CoIns between these states. 
We did not optimize singlet-triplet intersections, because ISC less likely occurs at localized points on the PES.\cite{mai___} Instead, we expect ISC to occur in extended regions of singlet-triplet interaction. It is interesting to note that the minima of all excited states under consideration and also both CoIns have $C_{2v}$ symmetry. 
Table~\ref{tab:excgeom} presents the optimized bond lengths and angles and corresponding energies. Comparison with other studies reporting these values shows that our bond lengths and angles are in excellent agreement with the results of Xie et al.\cite{xie_2013_JCP_14305} and with the MRCI values of L\'ev\^eque et al.\cite{leveque_2013_JCP_11}, while they differ slightly from the CCSD\cite{leveque_2013_JCP_11} and MRPT2\cite{li_2006_SCSB_289} results and the available experimental values (bonds are systematically longer). 
The energies of the stationary points show more variation than the geometries. Our calculated energies are systematically larger than the MRCI values of Xie et al.\cite{xie_2013_JCP_14305} and L\'ev\^eque et al.\cite{leveque_2013_JCP_11} but agree much more with the CCSD\cite{leveque_2013_JCP_11} and MRPT2\cite{li_2006_SCSB_289} values. We also note that our energies are slightly closer to the experimental results than those obtained by other MRCI calculations.\cite{xie_2013_JCP_14305,leveque_2013_JCP_11}

\begin{table}
  \centering
  \caption{Minimum energy geometries (all in $C_{2v}$) and CoIns of the excited states under consideration. $^1$CoIn denotes the CoIn between the singlet $^1A_2$ and $^1B_1$ states, $^3$CoIn the one between $^3A_2$ and $^3B_1$.}
  \label{tab:excgeom}
  \begin{tabular}{lcccccc}
    \hline
    State       &This work      &MRCI\cite{xie_2013_JCP_14305}    &MRCI\cite{leveque_2013_JCP_11}     &CCSD\cite{leveque_2013_JCP_11}     &MRPT2\cite{li_2006_SCSB_289}  &Exp.\\
    \hline
    \multicolumn{7}{c}{--- $r_{\text{SO}}$ (\AA) ---}\\
    $^1B_1$             &1.549  &1.550  &1.544  &1.515  &1.527  &\\
    $^1A_2$             &1.556  &1.558  &1.554  &1.527  &1.537  &1.53\cite{hamada_1974_CJP_1443}\\
    $^1$CoIn            &1.550  &1.550  &1.55   &1.52   &       &\\
    $^3B_1$             &1.517  &1.517  &       &       &1.484  &1.493\cite{brand_1971_JMS_616}\\
    $^3B_2$             &1.576  &       &       &       &1.561  &\\
    $^3A_2$             &1.556  &1.547  &       &       &1.535  &1.55\cite{heicklen_1980_RCI_315}\\
    $^3$CoIn            &1.547  &       &       &       &       &\\
    \hline
    \multicolumn{7}{c}{--- $\theta_{\text{OSO}}$ ($^\circ$) ---}\\
    $^1B_1$             &117.8  &118.7  &118.8  &120.7  &120    &\\
    $^1A_2$             &95.3   &93.6   &94.7   &93.4   &93     &99\cite{hamada_1974_CJP_1443}\\
    $^1$CoIn            &117.3  &115.8  &115.7  &114.2  &       &\\
    $^3B_1$             &124.7  &125.0  &       &       &129    &126.2\cite{brand_1971_JMS_616}\\
    $^3B_2$             &106.6  &       &       &       &106    &\\
    $^3A_2$             &95.0   &92.1   &       &       &94     &97\cite{heicklen_1980_RCI_315}\\
    $^3$CoIn            &111.1  &       &       &       &       &\\
    \hline
    \multicolumn{7}{c}{--- $E_{\text{adiab}}$ (eV) ---}\\
    $^1B_1$             &4.03   &4.1    &3.78   &4.03   &4.1    &3.96\cite{hamada_1975_CJP_2555}\\
    $^1A_2$             &3.55   &3.26   &3.32   &3.62   &3.58   &3.46\cite{hamada_1974_CJP_1443}\\
    $^1$CoIn            &4.03   &       &(3.79)\footnote{Value obtained by interpolation.\cite{leveque_2013_JCP_11}} &       &       &\\
    $^3B_1$             &3.41   &3.06   &       &       &3.2    &3.19\cite{brand_1971_JMS_616}\\
    $^3B_2$             &3.54   &       &       &       &3.41   &\\
    $^3A_2$             &3.32   &3.05   &       &       &3.29   &3.22\cite{heicklen_1980_RCI_315}\\
    $^3$CoIn            &3.59   &       &       &       &       &\\
    \hline
  \end{tabular}
\end{table}


\subsection{Dynamics in the singlet-manifold}\label{ssec:koeppel}

\begin{figure}
  \includegraphics[width=0.45\textwidth]{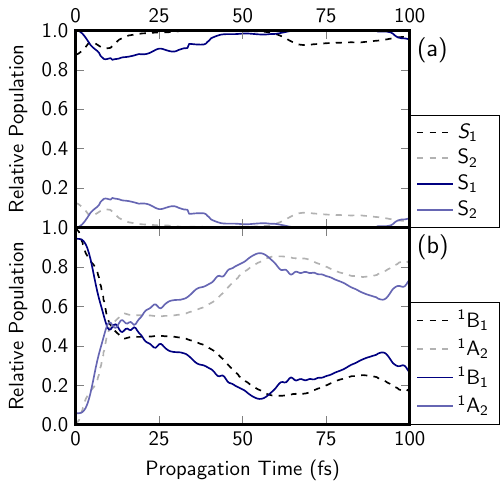}%
  \caption{Time evolution of the state populations from the QD simulations of L\'ev\^eque et al.\cite{leveque_2013_JCP_11} (dashed lines) and from the \textsc{Sharc} simulations of the present work (solid lines). In (a) MCH populations (``adiabatic'' in Ref.~\onlinecite{leveque_2013_JCP_11}), in (b) the spectroscopic populations (``diabatic'') of the $^1B_1$ and $^1A_2$ states. \label{fig:Population_SOnly}}%
\end{figure}

In order to further validate \textsc{Sharc}, we performed an MD simulation of an ensemble of 44 trajectories, including only the four lowest-lying singlet states $S_0$ to $S_3$. 
Since the $S_0$ and $S_3$ states are energetically separated from the $S_1$ and $S_2$ states and do not interact with them, the dynamics is nevertheless confined to $S_1$ and $S_2$. Thus, the results are directly comparable to those obtained by L\'ev\^eque et al.\cite{leveque_2013_JCP_11} who performed full-dimensional QD calculations on the coupled $S_1/S_2$ system. 
Figure~\ref{fig:Population_SOnly} shows the time-dependent populations in the MCH (panel (a)) and the spectroscopic representation (panel (b)). The results of L\'ev\^eque et al. are given by dashed lines, whereas our populations are given by solid lines. 

In the MCH picture, the adiabatic behaviour of the system is easily visible. In both the QD and the MD simulations, at $t=0$ the $S_1$ is predominantly populated and for later times always stays above 80\%. During the first few fs, some brief population transfer between $S_1$ and $S_2$ occurs, while the system is still close to the CoIn. In our simulation, a small number of trajectories switched to the $S_2$ surface. During the remaining course of the simulation, the $S_2$ is completely depopulated. In the simulation of L\'ev\^eque et al., the $S_2$ is partly repopulated after 60~fs, while in our simulation this happens only after about 100~fs. Comparing both methods, a fairly good agreement is reached, even though some minor details differ. 

In the spectroscopic representation, the $^1B_1$ state is initially populated, since it is the bright state. During the first 10~fs, a large fraction of the population is transferred to the $^1A_2$ state. The $^1A_2$ population reaches a maximum at about 60~fs, coinciding with $r_{\text{asym}}$ becoming close to zero after the first half-oscillation. Interestingly, in this representation, the populations of L\'ev\^eque et al. and of the \textsc{Sharc} simulation seem to be in closer agreement than in the MCH representation.


\subsection{Dynamics in the singlet-triplet manifold}\label{ssec:dynamics}

\begin{figure}
  \includegraphics[width=0.45\textwidth]{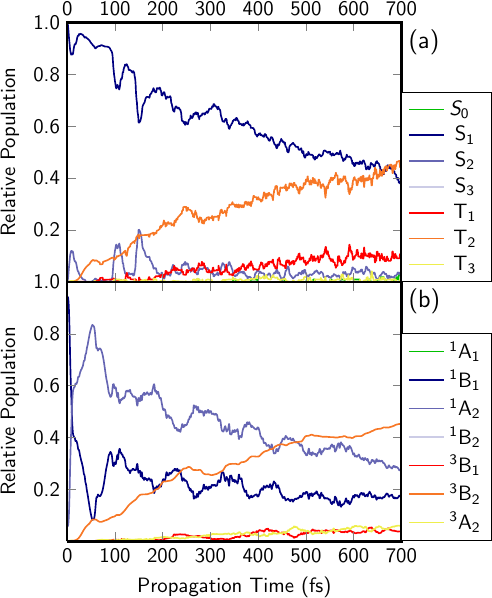}%
  \caption{Time evolution of the singlet and triplet state populations in the MCH (a) and the spectroscopic representation (b). \label{fig:Population_ST}}%
\end{figure}

Encouraged by the general agreement found in the dynamics performed within the singlet-manifold, an ensemble of 111~trajectories (all starting in $S_1$) was propagated for 700~fs, now allowing the interaction with triplet states. 

\begin{figure*}
  \includegraphics[width=\textwidth]{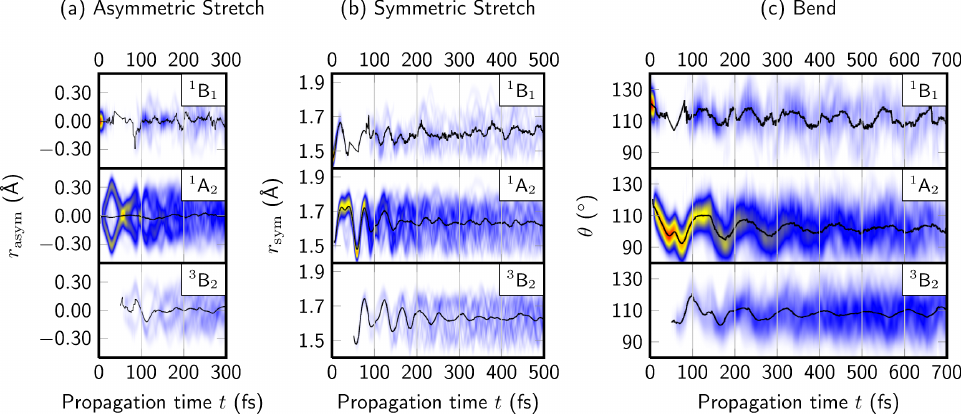}%
  \caption{Evolution of the internal coordinates of the ensemble; in (a) the asymmetric stretch coordinate $r_{\text{asym}}=\frac{1}{2}(r_1-r_2)$, in (b) the symmetric stretch coordinate $r_{\text{asym}}=\frac{1}{2}(r_1+r_2)$ and in (c) the bond angle $\theta$. The black lines give the expectation value of the internal coordinate. Note that discernable coherent motion is lost the fastest in the asymmetric stretch mode and the slowest in the bending mode, hence different time ranges are shown in the panels. $^3A_2$ and $^3B_1$ are not shown due to their low population.\label{fig:Motion}}%
\end{figure*}

Figure~\ref{fig:Population_ST} displays the relative populations of all singlet and triplet states depending on time. For simplicity, the populations of the three $M_S$ components of each triplet state were summed up. 
Analoguously to Figure~\ref{fig:Population_SOnly}, in panel (a) the MCH populations are given, while panel (b) shows the spectroscopic populations. In the evolution of the MCH populations, several processes can be identified. As before, the population is initially excited to the $S_1$ surface. Shortly after, some population is briefly transferred to the above-lying $S_2$ state, but the $S_2$ population quickly drops to zero within the first 50~fs, very similar to the singlet-only case. The $S_2$ surface is visited again at later times (most prominently around 100~fs and 150~fs), when the ensemble returns to the $S_1-S_2$ interaction region, but the $S_2$ population is negligible for other times.
Most importantly, significant ISC from $S_1$ to $T_2$ takes place. Interestingly, the $T_2$ population stays large during the simulation and does not quickly decay to the $T_1$ surface. We observed only a small number of $T_2-T_1$ transitions, accounting for about half of the $T_1$ population. The other half originated from ISC from $S_1$ to $T_1$.
The $T_3$ surface shows a very small population during the whole simulation time.

In the spectroscopic picture (panel (b)), the system is initially in the $^1B_1$ state. However, as in the singlet-only dynamics, within 10~fs, the character quickly changes to predominant $^1A_2$ character, as the ensemble moves around the CoIn towards smaller bond angles and towards $r_{\text{asym}}\neq0$. For later times, the system shows a damped oscillation of population between the two singlet states, with an oscillation period of 100-130~fs and an estimated damping constant of approximately 400~fs. Comparing with the MCH populations, it is notable that the $T_2$ population is very similar to the $^3B_2$ population. The populations of $^3B_1$ and $^3A_2$ are very small and their sum approximately corresponds to the $T_1$ population. A small oscillation between the $^3B_1$ and $^3A_2$ populations can be vaguely discerned.

A monoexponential function of the form
\begin{equation}
  f(t)=c\cdot(1-\text{e}^{-\frac{t}{\tau}})
\end{equation}
has been fitted to the triplet populations. Fitting the $^3B_2$ population alone, an ISC constant of 410~fs was obtained. Since the $^3B_1$ and $^3A_2$ states are not strongly populated and show the mentioned oscillation between each other, no separate time constant could be extracted for these states.
A fit of the sum of all triplet state population gives an effective time constant of 540~fs.

Figure~\ref{fig:Population_ST} can be directly compared to Figure~8 (a) of the recent paper of Xie et al.\cite{xie_2013_JCP_14305} Generally, in both simulations the same main processes are observed, oscillation within the singlet $B_1/A_2$ system and ultrafast ISC. However, compared to Xie et al., in our results the singlet population transfer appears to have a smaller oscillation amplitude and a slightly smaller oscillation period. Also, the oscillation seems to be damped more strongly in our calculations. Additionally, we observe a stronger population transfer to the triplet states, but with a comparable time constant. 


Figure~\ref{fig:Motion} shows the motion of the ensemble in all internal coordinates for all states with significant population. In panel (a), the time evolution of $r_{\text{asym}}$ is given. Initially, all trajectories are located at $r_{\text{asym}}\approx 0.0$ in the $^1B_1$ state. However, as soon as the simulation starts, the asymmetric stretch mode is strongly excited and the ensemble moves to values of $r_{\text{asym}}\neq 0.0$. The ensemble first returns to the initial  $r_{\text{asym}}\approx 0.0$ after about 50~fs, which is the half-cycle period of this mode. Also for 100~fs and 150~fs, recurrences can be observed. Different to the $^1A_2$ state, in the $^1B_1$ state the population is located at all times close to $r_{\text{asym}}= 0.0$. In the $^3B_2$ state, the excitation of this mode is as large as in the $^1A_2$ state.

In Figure~7 (b), the evolution of the symmetric stretch coordinate $r_{\text{sym}}$ is shown. The ensemble starts at $t=0$ with the very short bond lengths of the ground state equilibrium. After excitation a strong stretching of the \ce{S=O} bonds is observed. Initially, a large-amplitude, strongly anharmonic oscillation can be observed, reaching a maximal value of $r_{\text{sym}}\approx 1.73$~{\AA} after 21~fs and 38~fs. After 56~fs, the ensemble returns to the initial values for the first time. After about 200~fs, no clear oscillatory motion can be discerned anymore. 

The motion of the ensemble in the bending mode is depicted in Figure~7 (c). Starting from the ground state value of around 120$^\circ$, the system strongly decreases their bond angle upon excitation. After about 50~fs, a small increase of the bond angle is observed, which coincides with the first half-period of the asymmetric stretch motion. The oscillation period of the bending mode is 100-130~fs, and an in-phase motion of the ensemble can be discerned for the full 700~fs simulation time, showing 6 oscillation cycles. Most interestingly, these oscillation cycles nicely match the diabatic population transfer cycles depicted in Figure~\ref{fig:Population_ST}. A small average bond angle always coincides with a maximum of the $^1A_2$ population.


\begin{figure}
  \includegraphics[width=0.45\textwidth]{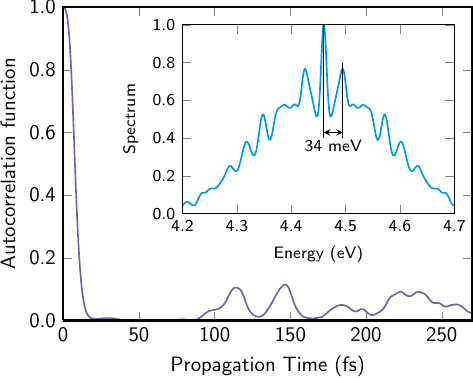}%
  \caption{Autocorrelation function obtained from Gaussian convolution of all trajectories. The inset shows the Fourier transform of the autocorrelation function (shifted by 4.46~eV) and the 34~meV vibrational spacing is indicated. \label{fig:AutoCorr}}%
\end{figure}

Figure~\ref{fig:AutoCorr} shows an approximate autocorrelation function, which was obtained from a Gaussian convolution of all trajectories within the internal coordinate space. For an explanation of the convolution scheme see the appendix. The FWHM of the Gaussians was taken as 0.1~\AA\ for $r_1$ and $r_2$ and 2$^\circ$ for $\theta$. The autocorrelation function shows a rapid drop during the first 20~fs, when the ensemble leaves the Franck-Condon region. The autocorrelation function shows a very small recurrence at 35~fs, which is caused by trajectories staying in the $^1B_1$ state instead of moving around the CoIn and towards the $^1A_2$ minima. The two most prominent recurrences are observed after 120~fs and 150~fs, arising from the ensemble returning to the Franck-Condon region after one oscillations in the bending mode and three half-oscillations in the asymmetric stretch mode. The Figure also shows the Fourier transform of the autocorrelation function (shifted by 4.46~eV, the vertical excitation energy to the $^1B_1$ state from table~\ref{tab:vertexc}), showing a 34~meV energy separation in the vibrational spectrum. Note that the convolution procedure did not include the electronic phase and thus the autocorrelation function is real-valued. Consequently, the spectrum is symmetric.


\section{Discussion}\label{sec:discussion}


\subsection{Spectrum}\label{ssec:spectrum_dis}

The vertical excitation energies in table~\ref{tab:vertexc} show a general agreement with the values obtained by other studies. While Xie et al.\cite{xie_2013_JCP_14305} and L\'ev\^eque et al.\cite{leveque_2013_JCP_11} both shifted their PESs to higher energies in order to agree with the experiment, our results are even slightly too high in energy compared to the experiment. This is apparent from the absorption spectrum in Figure~\ref{fig:Spectrum}, which is shifted to the blue relative to the experimental spectrum. This difference may be attributed to the double-zeta quality of the basis set. 
Despite its lesser accuracy, the double-zeta basis set was employed for the sake of computational efficiency during the MD simulations. 

The $S_0\rightarrow S_1$ transition completely dominates the spectrum (see Figure~\ref{fig:Spectrum}, while the $S_0\rightarrow S_2$ transition is very weak. Also in the work of L\'ev\^eque et al.\cite{leveque_2013_JCP_11} the $S_1$ dominates the spectrum, albeit to a lesser extent. The $S_1/S_2$ ratio is influenced by the location of the initial ensemble of geometries relative to the CoIn, with the bond angle being the most important coordinate in this respect. Slight displacements due to the level of theory can lead to notable differences in the initial population of $S_1$ and $S_2$. However, for the dynamics simulations this differences do not play a role (see section~\ref{ssec:koeppel} and Figure~\ref{fig:Population_SOnly}).

It is obvious that the vibrational structure of the spectrum is not reproduced by our semiclassical simulation. A proper simulation of the total absorption spectrum goes beyond the scope of this work, since it requires a quantum-mechanical treatment of the nuclear motion. A semi-classical simulation cannot account for quantization of the vibrational energy levels and also is not able to deliver nuclear wavefunction overlaps, which are necessary to obtain Franck-Condon factors. However, the recurrences observed in our approximate autocorrelation function correspond to an energy spacing of about 34~meV (see Figure~\ref{fig:AutoCorr}) or an oscillation period of 100-130~fs. This oscillation period corresponds to the recurrences of the wavepacket to the FC region of the initial photoexcitation. It is interesting that the beatings in the TRPEPICO measurements in paper~I\cite{wilkinson_2013_JCP_} of 145-155~fs compare reasonably well with the oscillation period in our simulations. Note that the signals observed in TRPEPICO arise from wavepacket recurrences to the FC region for ionization, which is possibly different from the FC region for the initial excitation. However, from our simulations we suggest that the recurrences to both FC regions are connected to the bending motion of the system. Thus, the oscillation periods in the TRPEPICO measurements and from the absorption spectrum reflect the very same motion on the $S_1$ surface, which we observe in our simulations.


\subsection{Potential energy surfaces}\label{ssec:PES_dis}

As has been described in \ref{ssec:PES}, the Franck-Condon point is located very close to the $^1B_1/^1A_2$ CoIn, leading to IC processes immediately after excitation. 

The location of the singlet-triplet interaction regions (given in Figure~\ref{fig:PES_all}) strongly influences the ISC processes which have been found to occur in the system. 
The intersection of $S_1$ with $T_3$ circularly surrounds the $S_2$/$S_1$ CoIn and the Franck-Condon region. Therefore, all trajectories leaving the Franck-Condon region are forced to pass through this intersection. However, since the region where singlet and triplet are nearly degenerate is quite narrow, population transfer between these states can only occur for a very brief period of time. Consequently, the ISC efficiency is negligible.

Instead, it is the singlet-triplet interaction region at large values of $r_{\text{asym}}$ and small bond angles $\theta$ that favors ISC. This region constitutes the outer turning point of the asymmetric stretch mode (see Figure~\ref{fig:PES_all} (b)) and shows comparably flat potentials, so that the trajectories spend a long time in this region after excitation of the asymmetric stretch mode. Thus, the time where singlet and triplet states can interact is large, allowing for a noteworthy ISC yield.

The data presented in table~\ref{tab:excgeom} compares well to values reported in the literature. Because our calculated bond lengths and angles are in excellent agreement with the ones reported by Xie et al.\cite{xie_2013_JCP_14305} and L\'ev\^eque et al.,\cite{leveque_2013_JCP_11} we can expect our excited-state PESs to be very similar to the ones used in the QD studies.\cite{xie_2013_JCP_14305,leveque_2013_JCP_11}

Interestingly, we obtain the lowest minimum energy for the $^3A_2$ instead of the $^3B_1$ state, which is reported as the lowest triplet state by other studies.\cite{li_2006_SCSB_289,brand_1971_JMS_616,heicklen_1980_RCI_315} This might influence the population dynamics in such a way that the $^3A_2$ would be on average too strongly populated. However, the $^3B_1/^3A_2$ system does not acquire a large population in the dynamics simulation, so that the above-mentioned discrepancy has no influence. 
The $^3B_2$ state was found to be the triplet state with the highest minimum energy, which is nevertheless still below both the $^1B_1$ and the $^1A_2$ minima. 
Furthermore, we note that we reproduce the difference of the minimum energies of the $^1A_2$ and $^1B_1$ states found in most other studies ($\approx 0.5$ eV).\cite{leveque_2013_JCP_11,li_2006_SCSB_289,hamada_1974_CJP_1443,hamada_1975_CJP_2555} 


\subsection{Dynamics in the singlet-manifold}\label{ssec:koppel_dis}

A good agreement is obtained between the QD simulation by L\'ev\^eque et al.\cite{leveque_2013_JCP_11} and our singlet-only MD simulations (recall Figure~\ref{fig:Population_SOnly}). In both studies, the wavepacket/ensemble basically moves all the time in $S_1$, despite having sufficient energy to reach the $S_2$. This behaviour can be attributed to the wavepacket/ensemble avoiding the CoIn because of the surrounding double-minimum potential (see Figure~\ref{fig:PES_all}). Following the potential energy gradient, the wavepacket/ensemble moves into regions of $r_{\text{asym}}\neq0$, where the two surfaces avoid each other and the non-adiabatic coupling is small. Additionally, population in the $S_2$ stays close to the CoIn because of the form of the $S_2$ potential, leading to rapid decay of $S_2$ population. Even though the $S_2$ is partly repopulated after the system returns to the FC region, the $S_2$ population remains very small. This is reasonable in the sense that the ensemble of trajectories already exhibits a considerable spread and the CoIn is missed by part of the ensemble.

Also for the diabatic populations (Figure~5 (b)), we found a nice agreement between the results of L\'ev\^eque et al.\cite{leveque_2013_JCP_11} and our work. During the first 10~fs, the $^1A_2$ population rises from zero to approximately 60\%, but increases much slower afterwards. This can be understood, realizing that in this timeframe the asymmetric stretch mode, which mixes the $^1B_1$ and the $^1A_2$ state, is excited, leading to comparable populations in both states. The diabatic $^1A_2$ population only rises to 90\% at the time when $r_{\text{asym}}$ gets close to zero after the first half-oscillation of the asymmetric stretch mode.

To sum up the findings so far, it is shown that the semi-classical dynamics gives results in satisfying agreement with the quantum dynamical results of L\'ev\^eque et al.\cite{leveque_2013_JCP_11}. Both show high population of $S_1$ at all times, with a fraction of the population oscillating between the two surfaces. In terms of diabatic states, an ultrafast, periodic transition between the bright $^1B_1$ state to the dark $^1A_2$ state was found. This periodic population transfer explains the spacings of about 30~meV and the FWHM of about 9~meV of the Clements bands. It also agrees with the intensity oscillations observed in the TRPEPICO spectra presented in paper~I.\cite{wilkinson_2013_JCP_}


\subsection{Dynamics in the singlet-triplet manifold}\label{ssec:dynamics_dis}

\begin{figure}
  \includegraphics[width=0.45\textwidth]{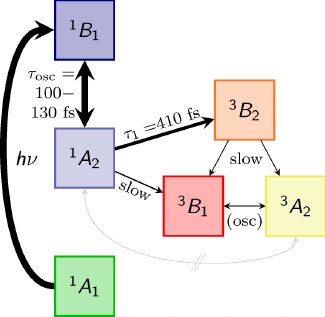}%
  \caption{Overview of the processes observed in the MD simulation of \ce{SO2}. Line width indicates the extent of population transfer. The time constants of the $^1B_1$ and $^1A_2$ transfer and the ISC processes is given. Population transfer between $^1A_2$ and $^3A_2$ is El-Sayed-forbidden, while population transfer between $^3B_2$ and the other triplet states is symmetry forbidden. \label{fig:overview}}%
\end{figure}

One of the most important results of this study is that ISC plays a significant role in the deactivation of \ce{SO2} and, as can be seen from Figure~\ref{fig:Population_ST}, even competes with IC on a timescale of hundreds of femtoseconds. In this regard, we agree with the findings of Xie et al.\cite{xie_2013_JCP_14305} who also observed ultrafast ISC in \ce{SO2} within a simplified model.
The electronic processes behind the time evolution of the populations plotted in Figure~\ref{fig:Population_ST} are summarized graphically in Figure~\ref{fig:overview}. After the initial photoexcitation from the $^1A_1$ state to the $^1B_1$ state, population is transferred adiabatically from $^1B_1$ to $^1A_2$, while staying on the $S_1$ surface. For later times, an oscillation of the population between these two states is observed, coinciding with the motion in the bending mode. This oscillation between $^1A_2$ and $^1B_1$  is the most important interaction occuring in this excitation region of the spectrum and according to L\'ev\^eque et al.\cite{leveque_2013_JCP_11} already accounts for the basic features of the absorption spectrum. The oscillations are primarily related to the bending motion of the molecule, since the two states cross along the bending coordinate. In the dynamics, the $S_3$ and the ground state do not interact with the populated states because of their large energetic separation. The absence of ground state relaxation was expected, since the $^1B_1/^1A_2$ system is known to be long-lived, with $\mu$s lifetimes being reported in the literature.\cite{greenough_1961_JACS_555,strickler_1968_JCP_1947} 

In our simulations, we additionally observe ISC from the $^1A_2$ to the $^3B_2$ and, to a smaller extent, to $^3B_1$. Almost 50\% of the total excited-state population is transferred via ISC to the $^3B_2$ state within 700~fs. This process is induced by strong elongation of the \ce{S=O} bonds (and small angles $\theta$), leading to a near-degeneracy of the singlet and triplet states, which then interact notably. The ISC to the $^3B_1$ state is less important because this state is close to the singlet surfaces only in the circular region surrounding the CoIn, where the system stays only for a very brief time (see section~\ref{ssec:PES_dis}). The $^3A_2$ state does not significantly participate in the ISC processes from $^1A_2$, since the transition is El-Sayed forbidden\cite{el-sayed_1963_JCP_2834} and thus the SOC matrix elements are very small (below 5 cm$^{-1}$). Additionally, the $^1A_2$ and $^3A_2$ surfaces are nearly parallel and do not show regions of small energy separation to facilitate ISC. 
IC within the triplet manifold is mainly restricted to the $^3B_1/^3A_2$ pair, which shows weak population transfer between each other. Based on the fact that the bending mode also leads to the crossing of $^3B_1$ and $^3A_2$, it is tempting to assume an oscillatory population exchange as in the singlet manifold. However, because of the low population of $^3B_1$ and $^3A_2$, this oscillation is not observed clearly. Since $^3B_2$ only interacts with $^3B_1$ and $^3A_2$ via triplet-triplet-SOC (about 55 cm$^{-1}$), IC from the $^3B_2$ state to $^3B_1$ or $^3B_1$ is slow. According to our findings, IC from $^3B_2$ and ISC from $^1A_2$ to the $^3B_1/^3A_2$ manifold are comparable in rate.

When comparing the results of our simulations to the work of Xie et al.,\cite{xie_2013_JCP_14305} some aspects of the dynamics nicely agree with each other, even though their study did not include the triplet states $^3A_2$ and $^3B_2$. First, in both simulations an oscillation of the system between the singlet states $^1A_2$ and $^1B_1$ has been found. This similar behaviour was expected, since the neglection of some triplet states should not influence the dynamics within the singlet manifold prior to ISC. The deviations in the oscillation times and damping behaviour between our work and the one of Xie et al.\ may be attributed to the different quantum chemical level of theory. 
Also, the general nuclear motion in both studies is comparable. This is reasonable in the sense that the PESs of the triplet state $^3B_2$ has a very similar shape as the $^3B_1$ state and thus neglecting the $^3B_2$ does not strongly influence the vibrational motion of the system.
However, in order to obtain correct ISC rates and pathways, the inclusion of all triplet states is mandatory, which is reflected in the different extent of ISC observed in our work and that of Xie et al.
While they find slightly below 30\% triplet population after 700~fs, we find already 50\% after this amount of time. The discrepancy can be explained by the fact that we include all relevant triplet states in the energetic region of the Clements bands. 
In particular, our simulations show that ISC leads to a predominant population of the $^3B_2$ and not of the $^3B_1$ as observed by Xie et al. Inspecting the PESs shown in Figure~\ref{fig:PES_all} in detail, it is found that both the $^3B_1$ and $^3B_2$ surfaces approach the singlet PES for long bond lengths; however, the energy gap between the $^1A_2$ and the $^3B_2$ is systematically smaller than $^1A_2-^3B_1$ gap. Thus, our simulations provide a clear indication that the inclusion of the $^3B_2$ state is very important for the excited-state dynamics. 

Our results also agree nicely with the experimental evidence of paper~I.\cite{wilkinson_2013_JCP_} There, a transient with an oscillation period of 155~fs has been measured. This period compares well with the 100-130~fs population oscillation found in our simulations for the $^1B_1/^1A_2$ pair.. Additionally, a damping of this population transfer with a time constant of a few hundreds of fs was found both in our simulation and in the experiment. Both oscillation period and damping constant agree well with the spacing and line width of the Clements bands.
The results reported in paper~I also strongly hint at the participation of ISC in the ultrafast dynamics of \ce{SO2}. First, it is suggested in paper~I that band~(2) (see Figure~7~(d) and (h) in paper~I) might be associated with ionization to quartet ion states based on ionization energetics. Since quartet ionic states could only be accessed from triplet neutral states, band (2) might be an indicator for ISC from singlet to triplet states.
A much stronger argument is the strong polarisation dependence of the photoelectron band (1) and its growth time components, with time scales between 150 and 750~fs. While the growth time component could be explained by intramolecular vibrational redistribution (IVR), in paper~I it is pointed out that the polarisation dependence  strongly suggests significant population transfer to $^3B_2$ via ISC, accessed from regions of the singlet PES with dominant $^1A_2$ character. This agrees very well with the observation of ISC to $^3B_2$ in our simulations. In summary, a very nice agreement both in the effective ISC time constants and the accessed states has been found between the experimental and theoretical works.

Nevertheless, a complete understanding of the \ce{SO2} TRPEPICO measurements is not possible based on the current simulations, because the ionization potentials and possibly photoionization yields need to be calculated along the trajectories.


\section{Conclusion}\label{sec:conclusion}

This work is -- to the best of our knowledge -- the first ab initio-based dynamics study to include all interactions between all singlet and triplet states present in the energy range of the first allowed band in the absorption spectrum of \ce{SO2}. 

It was found that upon excitation to the first allowed band system, the $^1B_1$ state is populated and subsequent nuclear motion leads the system periodically to a region of the PES with dominant $^1A_2$ character, all while adiabatically staying on the $S_1$ surface. The oscillation period of the interaction was found to be 100-130~fs and coincides with the nuclear motion in the bending mode. An oscillation with a similar period (145-155 fs) was reported in paper~I.\cite{wilkinson_2013_JCP_}
As was observed in the simulations and is supported by the TRPEPICO measurements of paper~I, ISC plays a significant role in the dynamics. With a time constant of 410~fs, the $^3B_2$ state is populated originating from regions of the PES with $^1A_2$ character, according to the El-Sayed rule.\cite{el-sayed_1963_JCP_2834} It was found that a strong elongation of one of the \ce{S-O} bonds and a small bond angle are prerequisites for this ISC process. Additionally, some population transfer to the $^3B_1$ state was observed, followed by interaction with the $^3A_2$ state via IC, similar to the $^1B_1/^1A_2$ interaction. These main processes have been summarized in Figure~\ref{fig:overview}.

In addition to a discussion in the light of the results of paper~I, our simulation were put into relation with two recent quantum dynamics studies on the energetic region of the Clements band system. L\'ev\^eque et al.\cite{leveque_2013_JCP_11} included in their dynamics simulations only the two singlet states $^1A_2$ and $^1B_1$. Within this two-state model, they already found the adiabatic dynamics on the $S_1$ surface and the oscillation between predominant $^1B_1$ and $^1A_2$ character which is one of the main processes observed in our simulations. 
In the second study, conducted by Xie et al.,\cite{xie_2013_JCP_14305} in addition to the two mentioned singlet states, the $^3B_1$ triplet state was included. Significant ISC originating from the $^1A_2$ surface to this triplet state was observed. However, our simulation showed that the inclusion of the two remaining triplets in this energetic region, the $^3A_2$ and in particular the $^3B_2$ state, is important for the correct description of ISC in this system. We found that $^3B_2$ is the primary triplet state populated by ISC.

The joint experimental and theoretical effort presented by paper~I and the current work shows how modern spectroscopic and \textsc{Sharc} dynamics coupled to multi-reference methods can shed light on intricate excited-state processes.


\appendix*
\section{Autocorrelation function}
In order to calculate the autocorrelation function based on the trajectories, first a density $\rho(t,r_1,r_2,\theta)$ was computed by Gaussian convolution. The calculation of the density is based on the discrete values $R_1(t,i)$, $R_2(t,i)$ and $\Theta(t,i)$, which are the \ce{S-O_1} and \ce{S-O_2} bond lengths and bond angle of trajectory $i$ at time $t$.
\begin{equation}
  \rho(t,r_1,r_2,\theta)=\sum\limits_{i=1}^{N_{\mathrm{traj}}}
  \mathrm{e}^{-\frac{(r_1-R_1(t,i))^2}{2c_1^2}}
  \mathrm{e}^{-\frac{(r_2-R_2(t,i))^2}{2c_1^2}}
  \mathrm{e}^{-\frac{(\theta-\Theta(t,i))^2}{2c_2^2}}
\end{equation}
The parameters $c_1$ and $c_2$ describe the FWHM of the Gaussian assigned to each trajectory. The width of the Gaussian was taken to be constant for all times $t$, and values of $c_1$=0.1\AA\ and $c_2$=2$^\circ$ were chosen.

Subsequently, the autocorrelation function was calculated from the overlap of the density at $t_0$ and at $t$:
\begin{equation}
  f(t)=\iiint \rho(t_0,r_1,r_2,\theta)\rho(t,r_1,r_2,\theta)\,\mathrm{d}r_1\,\mathrm{d}r_2\,\mathrm{d}\theta
\end{equation}
The spectrum in Figure~\ref{fig:AutoCorr} was obtained from Fourier-transformation of $f(t)$ and shifting by 4.46 eV, which is the vertical excitation energy from $^1A_1$ to $^1B_1$.


\begin{acknowledgments}
We deeply thank Horst K\"oppel and our collaboration partners Iain Wilkinson, Albert Stolow and Serguei Patchkowskii for many fruitful discussions. Financial support by the Austrian Science Fund (FWF): P25827-N28 and the COST action CM1204 (XLIC) as well as generous allocation of computer time at the Vienna Scientific Cluster (VSC2) are gratefully acknowledged.\label{sec:acknowledgments}
\end{acknowledgments}


%


\end{document}